\documentclass[%
 aip,
 jcp,
 amsmath,amssymb,
 reprint,%
]{revtex4-1}

\usepackage{graphicx}% Include figure files
\usepackage{overpic}% For \begin{overpic}...
\usepackage{dcolumn}% Align table columns on decimal point
\usepackage{bm}% bold math

\usepackage[utf8]{inputenc}
\usepackage[T1]{fontenc}
\usepackage{mathptmx}
\usepackage{etoolbox}
\usepackage{hyperref}

\newenvironment{subfigure}[2][]{\begin{minipage}[t]{#2}\centering}{\end{minipage}}

\makeatletter
\def\@email#1#2{%
 \endgroup
 \patchcmd{\titleblock@produce}
  {\frontmatter@RRAPformat}
  {\frontmatter@RRAPformat{\produce@RRAP{*#1\href{mailto:#2}{#2}}}\frontmatter@RRAPformat}
  {}{}
}%
\makeatother

\begin{document}

\preprint{JCP/000-XXX}

\title{Collective Variable-Guided Engineering of the Free-Energy Surface of a Small Peptide}
% Shifting the Thermodynamic Stability of Peptide Metastable States via Collective Variables
% Engineering Free-Energy Differences Between Metstable States in a Small Peptide Using Collective Variables
% Collective Variable–Guided Engineering of the Free-Energy Landscape Between Metastable States of a Small Peptide
% Shifting the Thermodynamic Stability of Peptide Metastable States via Collective Variables

\author{Muralika Medaparambath}
\affiliation{The Wolfson Department of Chemical Engineering, Technion – Israel Institute of Technology, Haifa 32000, Israel}
\affiliation{Faculty of Mathematics, Technion – Israel Institute of Technology, Haifa 32000, Israel}

\author{Alexander Zhilkin}
\affiliation{The Wolfson Department of Chemical Engineering, Technion – Israel Institute of Technology, Haifa 32000, Israel}

\author{Dan Mendels}
\affiliation{The Wolfson Department of Chemical Engineering, Technion – Israel Institute of Technology, Haifa 32000, Israel}

%\date{\today}

\begin{abstract}
Engineering the free-energy surfaces (FES) of proteins and peptides is central to controlling conformational ensembles and their responses to perturbations. However, predicting how chemical modifications such as point mutations reshape the FES and shift conformational equilibria remains challenging, particularly in data-scarce settings. Building on the Collective Variables for Free Energy Surface Tailoring (CV-FEST) framework, we develop a computational approach that leverages short, unbiased molecular dynamics trajectories to guide mutation analysis. Using the ten-residue $\beta$-hairpin CLN025 and a systematic library of its single-point mutants, we apply Harmonic Linear Discriminant Analysis (HLDA) to extract collective variables from the conformational data. We find that the HLDA eigenvector learned solely from short wild-type trajectories provides residue-level insight into the propensity of mutations at specific positions to thermodynamically stabilize or destabilize the folded state. Extending this analysis, we show that shifts in the leading HLDA eigenvalue across mutants, a measure of changes in separability between the conformational ensembles along the HLDA coordinate, correlate strongly with mutation-induced changes in the free-energy difference between states, as reflected in melting temperatures. Benchmarked against Replica Exchange Molecular Dynamics simulations, these findings suggest a promising and computationally affordable route toward guiding the engineering of biomolecular free-energy landscapes.
\end{abstract}

\maketitle

\section*{Introduction}

There has been remarkable progress in leveraging computational tools for protein engineering to accelerate research and development. This advancement has culminated in artificial intelligence (AI)-based structure prediction and generative models \cite{Jumper2021,Baek2021,Tsvi2024AF2PRL}, most notably AlphaFold, which have transformed biological research. Yet structure alone is not sufficient to explain protein function \cite{Cui2025}. Protein functionality is strongly influenced by thermodynamic and dynamic properties, or more broadly, by the features of the underlying free-energy surface (FES) and its response to perturbations such as mutations, co‑receptor‑induced structural changes, post‑translational modifications, or environmental factors like temperature and pH. This gives rise to a central challenge in biomolecular engineering: understanding and rationally modifying these FES, for example through targeted mutations to existing proteins.  Overcoming this barrier would unlock transformative applications, including designing proteins with tailored activities for targeted therapeutics \cite{Ebrahimi2023}, elucidating how mutant proteins contribute to cellular dysfunction and diseases, such as cancer \cite{CHITLURI2024}, and engineering protein- or peptide-based materials \cite{FinkelsteinZuta2024}.

To this end, a substantial body of experimental and computational work has focused on predicting how point mutations affect native-state protein stability, motivated in part by its relevance to protein engineering and disease interpretation  \cite{Pucci2022}. The emergence of Deep Mutational Scanning (DMS) \cite{Fowler2014} has greatly expanded the available pool of mutation–stability datasets, driving the development of numerous AI-based methods for predicting mutational effects on stability. Despite their growing popularity, however, these approaches often exhibit limited accuracy, owing to insufficient data in the target application domains as well as systematic biases and imbalances in the training datasets on which they rely \cite{Pucci2022,Huang2020,LipshSokolik2024,Sanavia2020,Zheng2024,Tsvi2024stabPRL}.

Physics-based approaches such as FoldX \cite{Guerois2002,Delgado2019} and Rosetta \cite{Kellogg2011,Goldenzweig2016,Leman2020} have proven useful in various high-throughput screening contexts via tools such as MutateX\cite{Papaleo2022MutateX} and RosettaDDGPrediction\cite{Papaleo2023RosettaDDGPrediction}.
However, depending on the application, these methods can also exhibit suboptimal accuracy, as highlighted in systematic assessments of protein stability and variant-effect predictors\cite{Sanavia2020,Gerasimavicius2022}, and, like ML-based approaches, they are primarily developed and evaluated for well-structured proteins, with much less established applicability to proteins with substantial flexibility or intrinsic disorder \cite{Fawzy2025}. Additionally, such approaches typically do not account for the contributions of unfolded states to protein stability \cite{Lee2026,Peccati2025}. Molecular dynamics (MD)–based methods, particularly enhanced sampling techniques such as replica-exchange molecular dynamics (REMD) \cite{Sugita1999}, metadynamics\cite{Laio2002}, and alchemical approaches such as thermodynamic integration\cite{Gapsys2020TI} have also been explored for this purpose. However, owing to their high computational cost and technical complexity, these methods are primarily applied to relatively small sets of carefully chosen systems and have not yet been widely adopted in high-throughput workflows.

To address these limitations, we explore in this work a novel approach to the engineering of protein and peptide FES, with particular emphasis on data-scarce regimes\cite{Salman&Sergey2025} in which large deep-learning (DL) models are often less effective. Our approach builds on the recently developed Collective Variables for Free Energy Surface Tailoring (CV-FEST) framework   \cite{MendelsDePablo2022,MendelsSciAdv2023}, which learns collective variables (CVs) from MD data to capture slow, relevant conformational transitions. CV-FEST is specifically suited to low-throughput settings, where only a limited number of FES calculations are feasible. 

As an initial proof of concept, we apply our methodology to the folding and unfolding of a short peptide. Peptides provide a convenient testbed: their modest size enables fully atomistic MD simulations across diverse sequences at reasonable computational cost, allowing detailed characterization of thermodynamic, dynamic, and kinetic behavior. At the same time, they capture key principles underlying protein stability and conformational dynamics. Beyond their methodological utility, peptides are of practical interest due to their central roles in biological recognition and regulation \cite{CUNNINGHAM2017}, their growing use as therapeutic modalities \cite{Wang2022,Gokhale2014,Tsomaia2015}, and their intrinsic conformational flexibility, which makes them valuable models for disordered and partially ordered protein systems \cite{Coskuner2019,Toprakcioglu2022}. Peptides also hold promise for antibacterial applications and responsive biomaterials \cite{FinkelsteinZuta2024}.

Using the ten-residue $\beta$-hairpin CLN025\cite{Honda2008} (Fig.~\ref{fig:cln025}(a)) as a representative model system, we test whether CV-FEST-derived CVs, learned from short unbiased MD simulations, can (i) identify mutation-sensitive positions in the peptide and (ii) predict shifts in the folding free-energy difference $\Delta G$ between folded and unfolded basins.

\section*{Methods}

The introduced methodology  builds on CV-FEST, a framework we previously developed for free energy engineering in systems dominated by rare events\cite{MendelsDePablo2022,MendelsSciAdv2023}. CV-FEST was developed to address several practical challenges associated with applying many machine-learning–based approaches to molecular and materials systems, including large data requirements, limited interpretability, and restricted generalizability beyond the training domain. Unlike conventional methods that depend on extensive datasets drawn from many simulations or experiments, which are often unavailable or strongly biased, CV-FEST leverages dynamical information obtained from MD trajectories of a single system or a small set of related systems. The framework is based on the assumption that the essential behavior of slow collective modes can be captured through a low-dimensional representation of the FES defined by an appropriate set of CVs. This compact representation concentrates the most relevant information about the system into a reduced design subspace of possible system modifications, which can be efficiently explored and optimized. \cite{MendelsDePablo2022,MendelsSciAdv2023}.

CVs constitute a central component of many enhanced sampling techniques \cite{Torrie1977,Laio2002,Darve2001}, guiding the exploration of the system's phase space and defining the reduced coordinate space in which the corresponding FES is constructed. CVs are functions of the atomic coordinates, $\mathbf{s}(\mathbf{R})$, and their probability distribution is given by
\begin{equation}
P(\mathbf{s'}) = \int d\mathbf{R}\delta\left[\mathbf{s'}-\mathbf{s}(\mathbf{R})\right] P(\mathbf{R}),
\end{equation}
where $P(\mathbf{s'})$ denotes the probability of observing a given CV value $\mathbf{s'}$, $P(\mathbf{R})$ is the Boltzmann probability distribution, and $\delta$ is the Dirac delta function. The associated FES is then defined as 
\begin{equation}
    F(\mathbf{s}) = -k_{\mathbf{B}}T\ln P(\mathbf{s}),
\end{equation}
where $k_{\mathbf{B}}$ is the Boltzmann constant and $T$ is the temperature. 

CVs that enable efficient sampling typically capture the essential physics of a system’s slow processes, making them valuable for studying structure-dynamics-function relationships. Although constructing such CVs is challenging as the relationship between molecular structure and the underlying FES tends to be complex , previous work\cite{Mendels2018,Piccini2018,Mendels2018_hlda,Bonati2020,Chen2018AutoencodersCV,Ribeiro2018RAVE,Noe2019BoltzmannGenerators,Wehmeyer2018TLAE} has shown that ML-based approaches can be effective. 

\paragraph*{Collective Variables:}
In the current study we construct the CVs using Harmonic Linear Discriminant Analysis (HLDA)\cite{Mendels2018,Piccini2018,Mendels2018_hlda,Bonati2020,Zhang2019,Rizzi2019,Brotzakis2019AugHLDA}. HLDA represents each CV as linear combination of user-defined descriptors, which facilitates interpretability: descriptors with larger weights contribute more strongly to the system’s slow collective behavior. In addition, training HLDA is straightforward and data-efficient, requiring only limited sampling within the metastable basins between which the relevant rare events occur. 

Mathematically, given a descriptor set, HLDA first estimates the mean vectors $\boldsymbol{\mu}$ and covariance matrices $\boldsymbol{\Sigma}_I$ of the descriptors for each metastable state $I$, using short unbiased trajectories of those states. It then seeks a direction $\mathbf{W}$ in the $N_d$-dimensional descriptor space that maximally separates the projected training distributions by maximizing the Rayleigh quotient
\begin{equation}
J(\mathbf{W}) = \frac{\mathbf{W}^T \mathbf{S}_b \mathbf{W}}{\mathbf{W}^T \mathbf{S}_w \mathbf{W}}.
\end{equation}
where for two metastable states A and B, the between-class scatter matrix is
\begin{equation}
\mathbf{S}_b = (\boldsymbol{\mu}_A - \boldsymbol{\mu}_B)(\boldsymbol{\mu}_A - \boldsymbol{\mu}_B)^T,
\end{equation}
while the within-class scatter matrix uses the HLDA harmonic average
\begin{equation}
\mathbf{S}_w = \left( \boldsymbol{\Sigma}_A^{-1} + \boldsymbol{\Sigma}_B^{-1} \right)^{-1},
\end{equation}
where $\boldsymbol{\mu}_{A,B}$ and $\boldsymbol{\Sigma}_{A,B}$ are the means and covariances of states A and B respectively. 

Under the normalization $\mathbf{W}^T \mathbf{S}_w \mathbf{W} = 1$, maximizing $J(\mathbf{W})$ is equivalent to solving the generalized eigenvalue problem
\begin{equation}
\mathbf{S}_w^{-1} \mathbf{S}_b \mathbf{W} = \lambda \mathbf{W}.
\end{equation}
The leading eigenvector defines the HLDA CV, the direction minimizing overlap between projected folded and unfolded ensembles, while the associated eigenvalue $\lambda$ quantifies separability along this axis (larger $\lambda$ indicating better discrimination), a metric we exploit herein.

\begin{figure}[t]
  \centering
  \begin{subfigure}[t]{0.26\columnwidth}
    \centering
    \begin{overpic}[width=\linewidth]{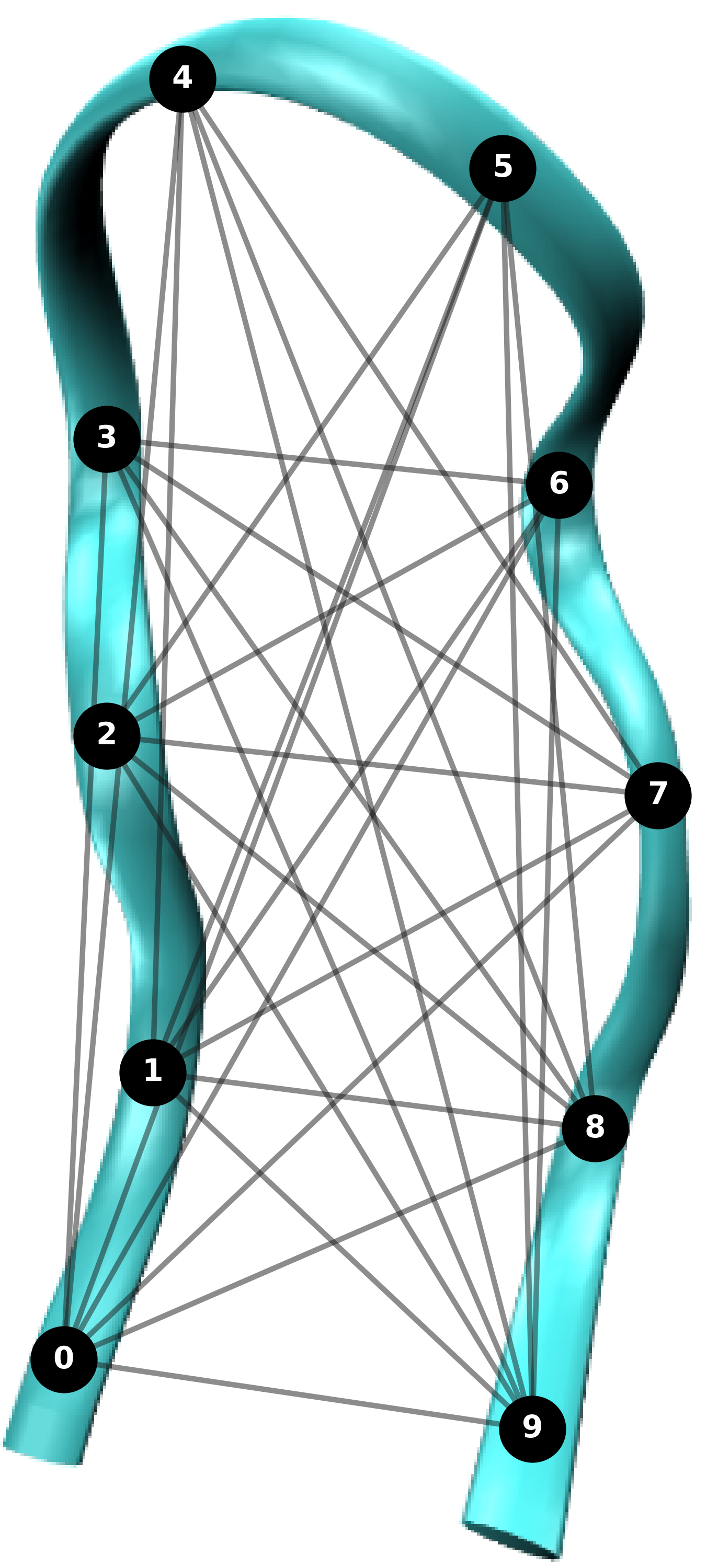}
      \put(2.5,100){\scriptsize\textbf{(a)}}
    \end{overpic}
    % \caption{CLN025 and the inter-residue backbone distance descriptors (orange lines).}
  \end{subfigure}
  \hfill
  \begin{subfigure}[t]{0.70\columnwidth}
    \centering
    \begin{overpic}[width=\linewidth]{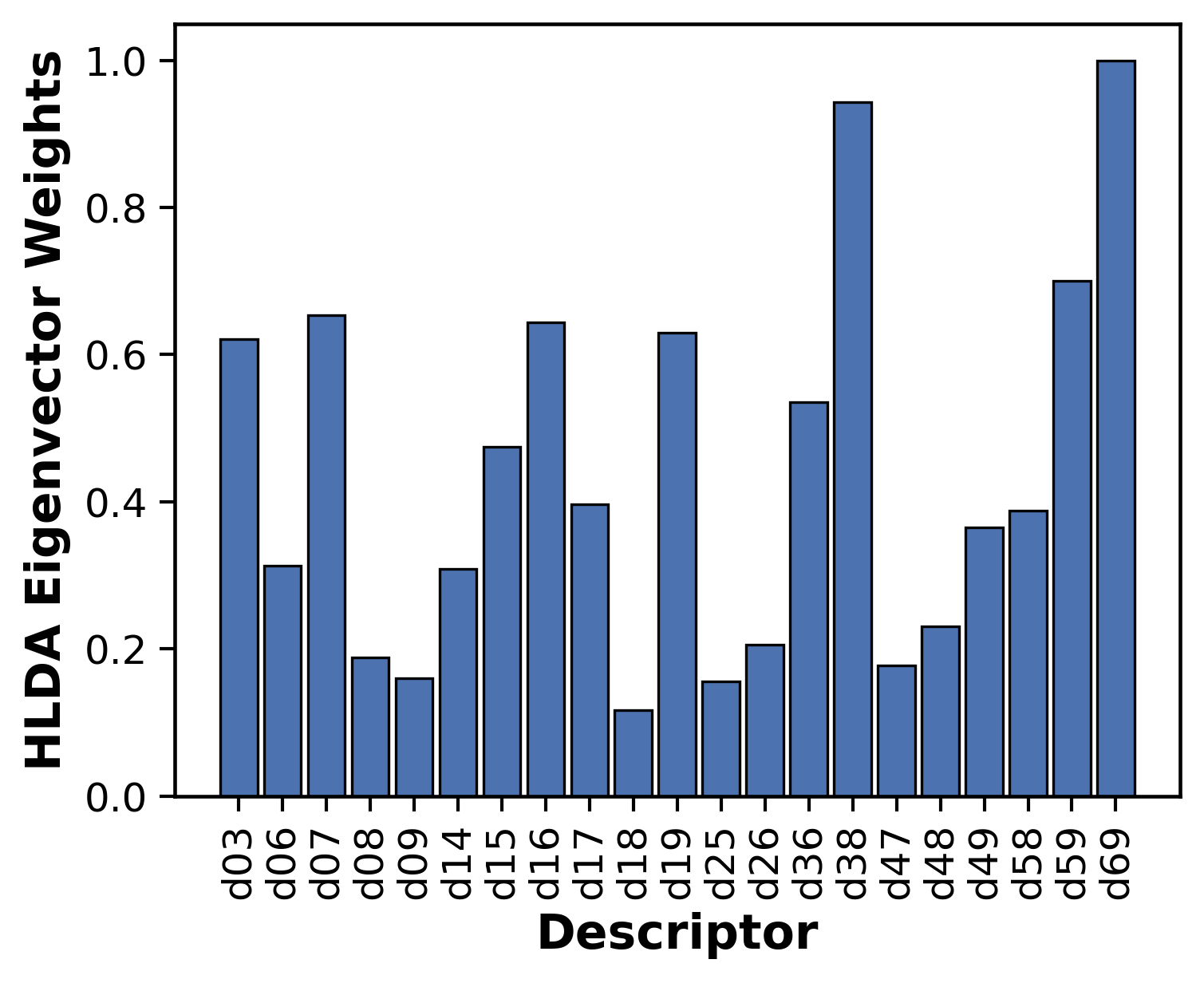}
      \put(2,82){\scriptsize\textbf{(b)}}
    \end{overpic}
    % \caption{Absolute HLDA eigenvector weights for each distance descriptor.}
  \end{subfigure}
  \caption{(a) Structural model of CLN025 with inter-residue distance descriptors (black lines). (b) Absolute HLDA eigenvector weights for each distance descriptor.}
  \label{fig:cln025}
  \end{figure}

These eigenvectors assign interpretable weights to the descriptors (Fig. \ref{fig:cln025}(b)), revealing their relative contributions to separability along the CV axis. In previous studies\cite{MendelsDePablo2022,MendelsSciAdv2023}, we showed that targeted modulation of forces associated with highly weighted descriptors enables systematic shifts in the relative thermodynamic stability of the states. Building on this concept, we extend the framework to a more realistic setting by introducing explicit chemical perturbations in the form of point mutations, rather than directly tuning interaction parameters. The resulting eigenvectors and eigenvalues are then used to guide mutation selection and to predict the associated changes in the free-energy difference $\Delta G$.

For CV construction, we used all inter-residue backbone distances, excluding nearest- and next-nearest-neighbor pairs, yielding a total of 28 descriptors  (Fig.~\ref{fig:cln025}(a)). To improve numerical stability in the HLDA calculations, we performed correlation-based pruning prior to HLDA by removing highly correlated descriptor pairs (see more details in the Computational Details). Residue-importance scores were computed from the wild-type (WT) HLDA eigenvector by aggregating descriptor-level contributions at the residue level. After obtaining the leading HLDA eigenvector $\mathbf{W}$  over the pruned descriptor set (with weights $w_{ij}$), descriptor weights were assigned to residues by grouping all descriptors involving each residue $r$. The per-residue importance was then defined as the mean absolute weight,
\begin{equation}
    I_r=\frac{1}{2N_r}\sum_{(i,j):\, i=r \,\vee\, j=r} |w_{ij}|,
\label{eq_res_imp}
\end{equation}

where the factor of $2$ is given that $w_{ij}=w_{ji}$ , and $N_r$ is the number of descriptors associated with residue $r$. Descriptors removed during pruning were included by assigning them the weight of the retained descriptor to which they were most strongly correlated.
% In addition, we used the leading HLDA eigenvalue, \$\lambda\$, as a measure of folded–unfolded separability along the HLDA CV axis. Mutation-induced changes in separability were quantified relative to the wild type through the eigenvalue shift, \$\Delta\textbackslash{}lambda=\textbackslash{}lambda\_{\textbackslash{}mathrm\{mut\}}-\textbackslash{}lambda\_{\textbackslash{}mathrm\{WT\}}\$, providing a compact measure of how different substitutions strengthen or weaken ensemble separation. 

As noted above, because HLDA assigns the largest weights to descriptors most strongly associated with the considered slow mode, modifying the forces linked to these descriptors, e.g., through chemical perturbations such as point mutations, is expected to have the greatest impact on the free-energy difference between the participating metastable states. Because these descriptors also dominate separability along the one-dimensional CV axis, we hypothesized that mutation-induced changes in separability could serve as a surrogate for alterations to the FES governing the rare event.  To test this, we use the leading HLDA eigenvalue $\lambda$  as a measure of folded–unfolded separability and quantify mutation-induced changes relative to the wild type via  $\Delta\lambda=\lambda_{\mathrm{mut}}-\lambda_{\mathrm{WT}}$, providing a compact metric of how substitutions strengthen or weaken ensemble separation and, by extension, alter the free-energy difference between the metastable basins.

\section*{Computational Details}
\paragraph*{Simulation Setup:}
All MD simulations were performed using GROMACS \cite{Abraham2015} patched with PLUMED \cite{Tribello2014}. For each system WT and mutants, the minimum-enthalpy structure was used as the initial configuration. Systems were modeled using the CHARMM22* force field with the TIP3P water model \cite{Piana2011,Jorgensen1983}, solvated in a cubic box containing approximately 1{,}800 water molecules and neutralized with sodium ions. Following energy minimization, systems were equilibrated in the canonical (NVT) ensemble at 340~K using the velocity-rescaling thermostat \cite{Bussi2007} with a 2~fs timestep. Covalent bonds involving hydrogen atoms were constrained using LINCS \cite{Bekker1998}, and long-range electrostatics were treated using the particle-mesh Ewald (PME) method \cite{Darden1993}.

\paragraph*{Training Data:}
For each system, two short unbiased trajectories were generated to provide training data for CV construction: one initiated from a folded, native-like structure and one initiated from an unfolded conformation obtained by temporarily restraining the peptide end-to-end distance prior to release. Each trajectory was propagated for 100~ns at 340~K and used for subsequent analysis, following preliminary assessments that confirmed convergence of the HLDA metrics.

In the short unbiased trajectories used to construct the HLDA training  ensembles, spontaneous transitions between folded and unfolded conformations were occasionally observed. To limit mixing between the state-specific ensembles, folded and unfolded states were defined using a root-mean-square deviation (RMSD) CV computed relative to a system-specific minimum-enthalpy reference structure. Because the resulting ensemble statistics showed some sensitivity to the chosen boundaries, we repeated the analysis over a range of RMSD threshold pairs. The thresholds yielding the strongest WT residue-importance correlations were practically identical to those producing the strongest $\Delta\lambda$--$\Delta T_m$ correlations, considering the former is evaluated on the WT alone whereas the latter spans both WT and mutants. Importantly, both correlations persist over broad ranges of threshold values (Figs. \ref{fig:wt_res_imp_heatmap_si} and~\ref{fig:threshold_heatmap}), with the upper limit of the folded state corresponding to the system’s free-energy barrier region (Fig.\ref{fig:melting_curves}(a)), and, interestingly, the unfolded-state boundary aligning with RMSD values characteristic of the fully unfolded ensemble while excluding partially folded conformations.

\paragraph*{Descriptor Set Preprocessing:}
To improve numerical stability in the HLDA calculations, we performed correlation-based pruning prior to HLDA by removing descriptor pairs with absolute correlation exceeding a tolerance $r_{\mathrm{tol}}$ (see more details in the Computational Details section). We carried out this pruning separately within the folded and unfolded ensembles using Pearson correlations, and then kept only the descriptors that survived both filters. Since eigenvector-derived HLDA metrics are more perturbation-sensitive than eigenvalues~\cite{Stewart1990}, we applied a stricter tolerance ($r_{\mathrm{tol}}=0.93$) for eigenvector-based analyses than for eigenvalue-based measures ($r_{\mathrm{tol}}=0.98$). This choice further improved the stability and robustness of the results with respect to the definition of the state boundaries.

\paragraph*{REMD Simulations and Melting-Temperature Estimations:}
REMD simulations \cite{Sugita1999} were performed in the NVT ensemble using 25 replicas. Temperatures were distributed geometrically with a scaling factor $a=1.0195$ starting from 340~K, and exchange attempts were carried out every 1~ps. Conformational states were classified using the RMSD to the native folded reference structure, with RMSD values defining the folded and unfolded basins in accordance with the computed FES (Fig.\ref{fig:fes_plots}(a)). The first 100~ns of each replica trajectory were discarded for equilibration. For each temperature, we computed the folded-state probability as $P_{\mathrm{folded}}(T)=N_F/(N_F+N_U)$, where $N_F$ and $N_U$ are the numbers of frames assigned to the folded and unfolded basins, respectively \cite{Rao2005}. The melting temperature $T_m$ was defined by $P_{\mathrm{folded}}(T_m)=0.5$. Each system was simulated with multiple independent REMD runs; within each run, trajectories were divided into blocks of varying sizes and $T_m$ was recomputed for each block to assess convergence. Final $T_m$ values were obtained by combining run-wise estimates, and uncertainties were estimated from the variability across blocks. Replica-exchange mixing and sampling convergence were assessed using standard diagnostics such as temperature-space random walks, RMSD time series for the lowest-temperature replica, and block-size convergence of $T_m$ (Fig.~\ref{fig:si_remd_convergence}), following the approach in Ref.~\cite{remd_convergence}.

\section*{Results}

To assess whether our framework can help guide the selection of point mutations for reshaping the WT FES, we analyzed a systematic mutation scan of the CLN025 peptide and compared CV–derived quantities with thermodynamic benchmarks obtained from REMD simulations. In total, we introduced 36 single-point mutations across seven of the ten residues of the peptide, selected to provide broad coverage along the sequence. At each site, the native residue was substituted with four to eight amino acids spanning distinct physicochemical classes, including charged, polar, hydrophobic, and aromatic residues, in order to evaluate their effects on the peptide’s FES.

\begin{figure}[htbp]
\centering
\begin{subfigure}[b]{0.49\columnwidth}
\centering
\vspace{0pt}
\begin{overpic}[width=\linewidth]{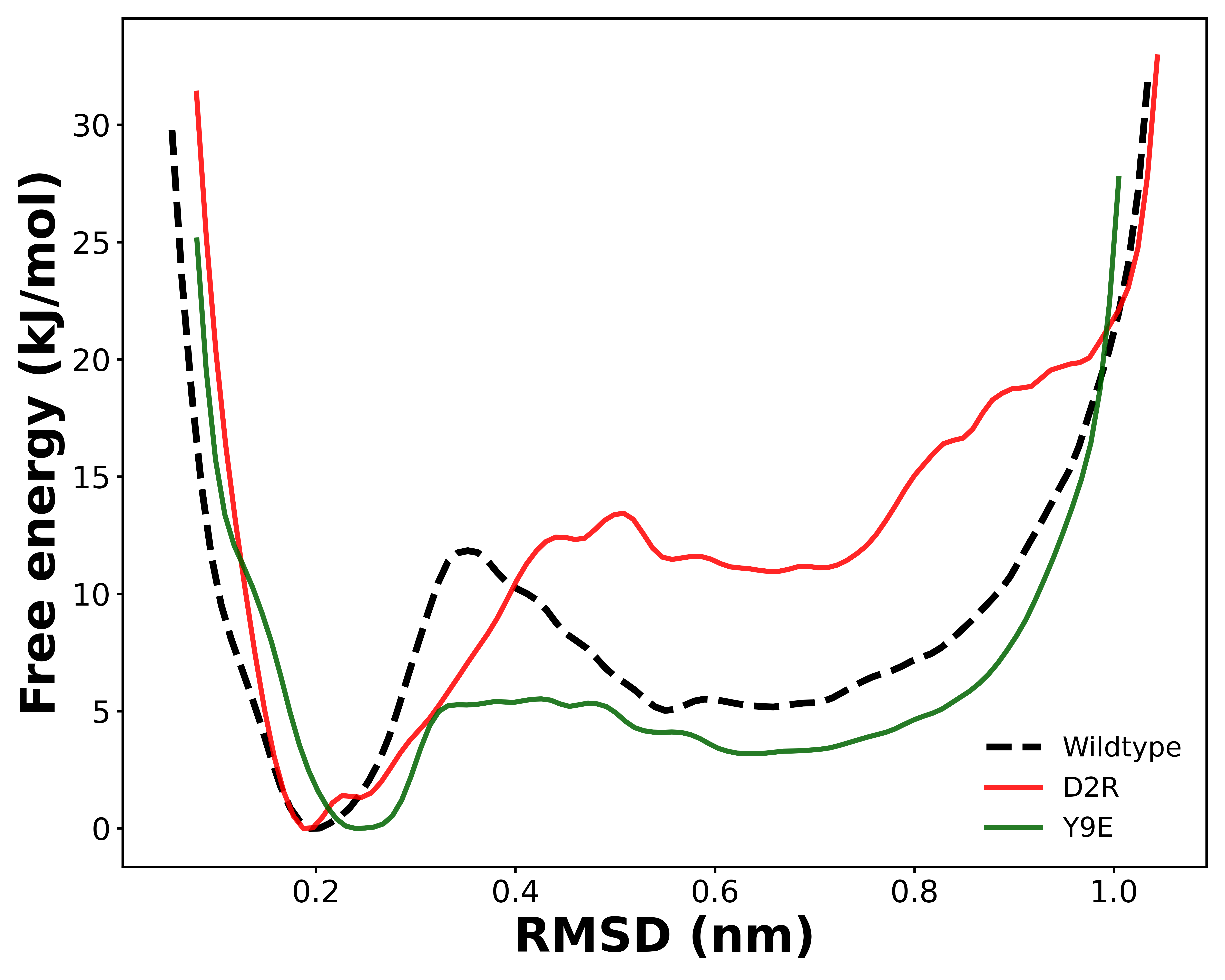}
  \put(2,82){\scriptsize\textbf{(a)}}
\end{overpic}
\label{fig:fes_plots}
\end{subfigure}
\hfill
\begin{subfigure}[b]{0.49\columnwidth}
\centering 
\vspace{0pt}
\begin{overpic}[width=\linewidth]{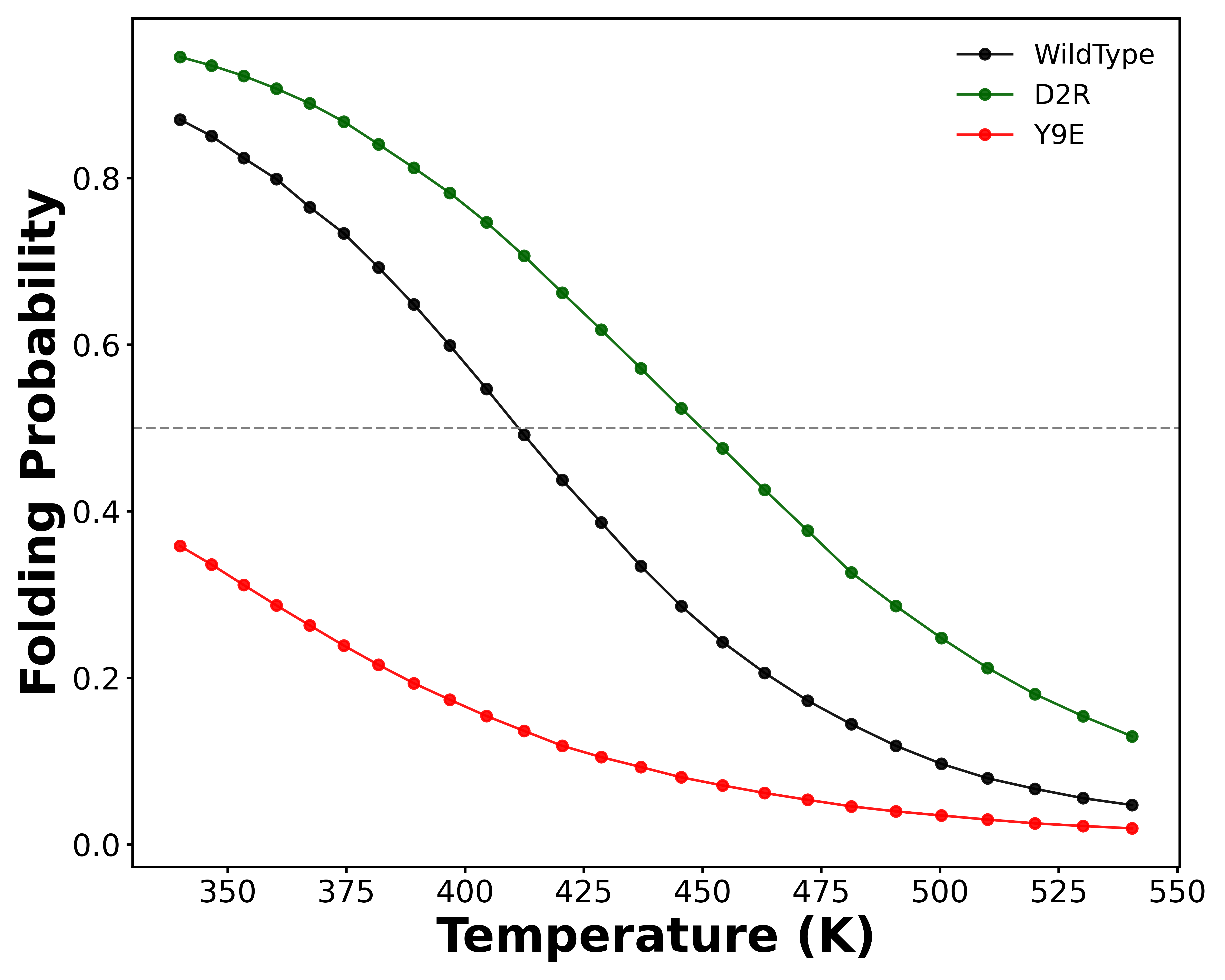}
  \put(2,82){\scriptsize\textbf{(b)}}
\end{overpic}
\label{fig:pvst}
\end{subfigure}
\caption{Comparison of WT CLN025 and two representative mutants:
(a) Free-energy profiles along the RMSD CV at 340~K showing mutation-dependent changes.
(b) Folded probability versus temperature from REMD simulations, with $T_m$ defined at $P_{\mathrm{folded}}=0.5$.}
\label{fig:melting_curves}
\end{figure}

Mutation-induced changes in the FES were characterized using REMD by examining free-energy profiles projected along the RMSD CV, which exhibit pronounced, mutation-dependent variations   (Fig.~\ref{fig:melting_curves}(a)). These variations manifest as shifts in the relative free-energy balance between folded and unfolded regions and lead to corresponding changes in the peptide's $T_m$, defined by $\Delta G(T_m)=0$ (equivalently $P_{\mathrm{folded}}(T_m)=0.5$), as illustrated by the temperature dependence of $P_{\mathrm{folded}}$ in Fig.~\ref{fig:melting_curves}(b)). To evaluate the ability of the HLDA CV to capture these mutation-induced FES changes, we compare the melting temperatures obtained from REMD with HLDA-CV-derived descriptors.

\subsection*{WT HLDA-derived residue-importance scores}
As a starting point we first examine whether an HLDA model trained on the WT folded and unfolded ensembles can provide residue-level guidance on mutation sensitivity. To this end, we computed a WT HLDA residue-importance score (Eq.\ref{eq_res_imp}) by aggregating the absolute values of the individual descriptor weights in the HLDA eigenvector (see Fig.~\ref{fig:cln025}(b)) associated with each residue, thereby quantifying how strongly each site is, on average, linked to the folded–unfolded transition. The residue-importance scores were then compared with the mean change in the peptide’s melting temperature, $\Delta T_m = T_{m}^{\mathrm{mut}} - T_{m}^{\mathrm{WT}}$, averaged over the different substitutions introduced at each site (i.e., across the set of amino acids used). 

\begin{figure}[t]
\centering
\begin{subfigure}[b]{0.49\columnwidth}
\centering
\vspace{0pt}
\begin{overpic}[width=\linewidth]{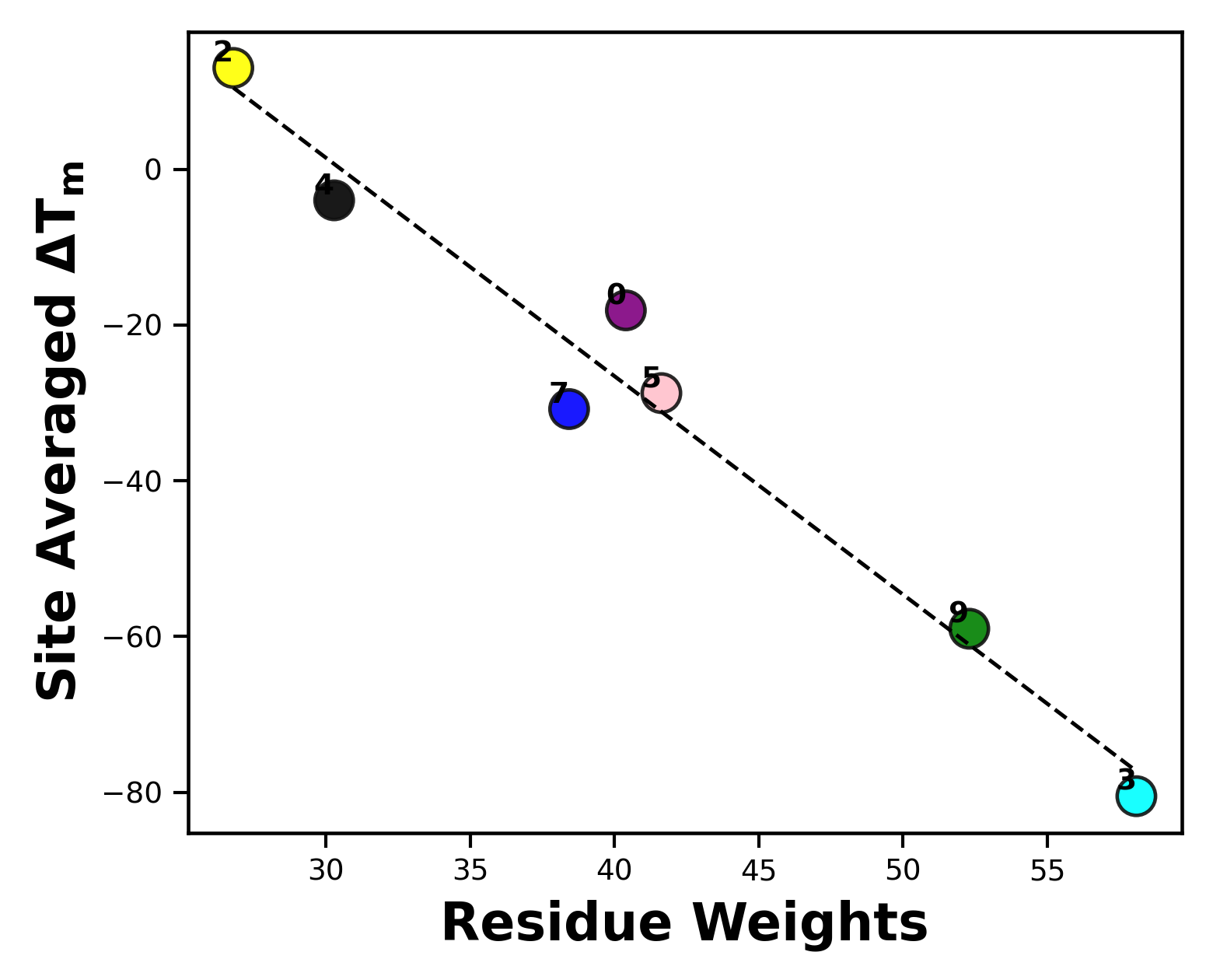}
  \put(2,82){\scriptsize\textbf{(a)}}
\end{overpic}
\label{fig:wt_res_weights_scatter}
\end{subfigure}
\hfill
\begin{subfigure}[b]{0.49\columnwidth}
\centering
\vspace{0pt}
\begin{overpic}[width=\linewidth]{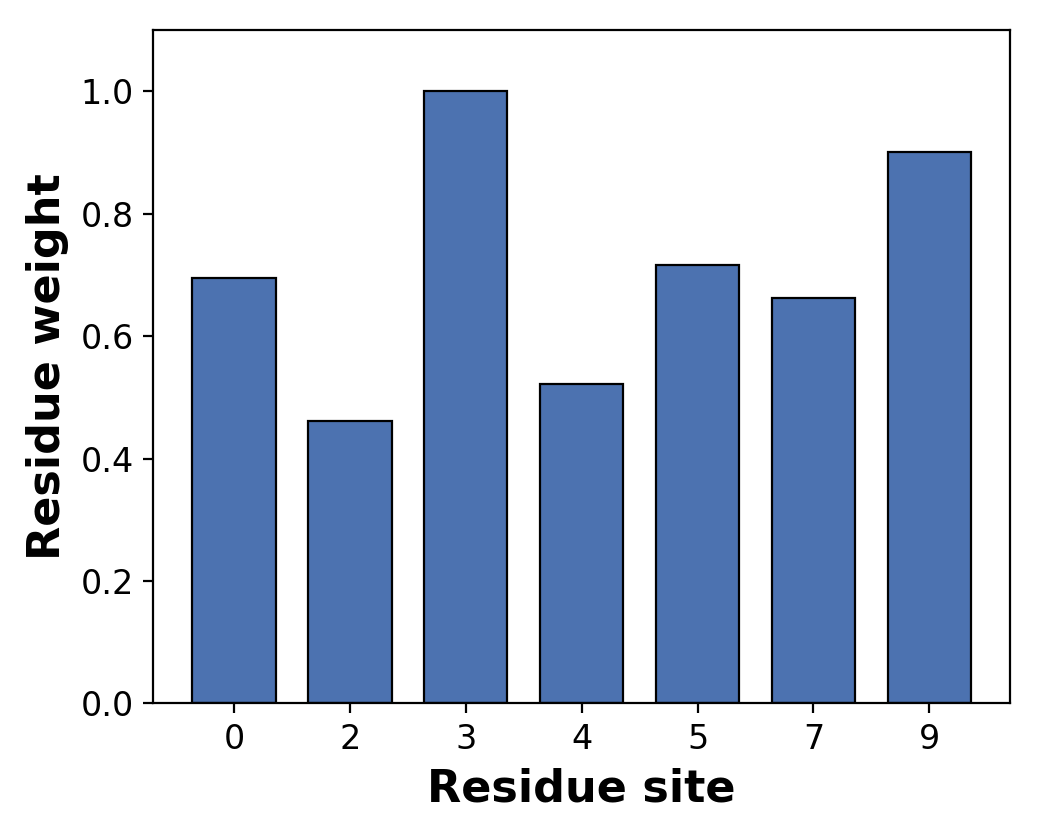}
  \put(2,82){\scriptsize\textbf{(b)}}
\end{overpic}
\label{fig:wt_res_weights_bar}
\end{subfigure}
\caption{WT HLDA eigenvector–derived residue weights versus per-residue mean changes in the peptide’s melting temperature: (a) Scatter plot summarizing the relationship between the WT residue-score computed using Eq.\ref{eq_res_imp} and the mean $\Delta T_m$ across mutations at each site (Pearson $r=-0.98$, $p=8.09\times10^{-5}$).
(b) Per-residue WT weights shown as bars to highlight differences across positions.}
\label{fig:wt_res_importance}
\end{figure}

We observe a pronounced inverse correlation between these quantities  (Fig.~\ref{fig:wt_res_importance}(a)). This trend suggests that residues that are more prominently represented on average in the HLDA CV tend to correspond to positions where substitutions are associated with larger destabilization of the folded state. Conversely, the residue with the lowest average HLDA weight was found to exhibit stabilizing substitutions. Across the full set of single mutants, destabilizing variants constitute the majority and dominate the HLDA signal, which is consistent with the general observation that stabilizing mutations are generally harder to capture accurately than destabilizing ones\cite{Zheng2024}. At the same time, the distribution of $\Delta T_m$ values across different substitutions at a given site indicates that, although certain positions appear more sensitive to perturbation, both the direction and magnitude of the response depend on the specific amino acid introduced. 

\subsection*{Mutation-specific stability changes}
We next move from site-averaged sensitivity to mutation-specific effects, asking whether the framework can help predict the $T_m$ shift induced by a particular substitution. As noted above, we hypothesized that mutation-induced changes in the separability between the folded and unfolded states along the HLDA CV could serve as a proxy for the corresponding change in the free-energy difference between the two states, $\Delta \Delta G$, relative to wild type. To test this idea, we recomputed the HLDA CV for each mutant peptide and measured the corresponding change in $\lambda$.

\begin{figure}[t]
\centering
\begin{subfigure}[b]{0.49\columnwidth}
\centering
\vspace{0pt}
\begin{overpic}[width=\linewidth]{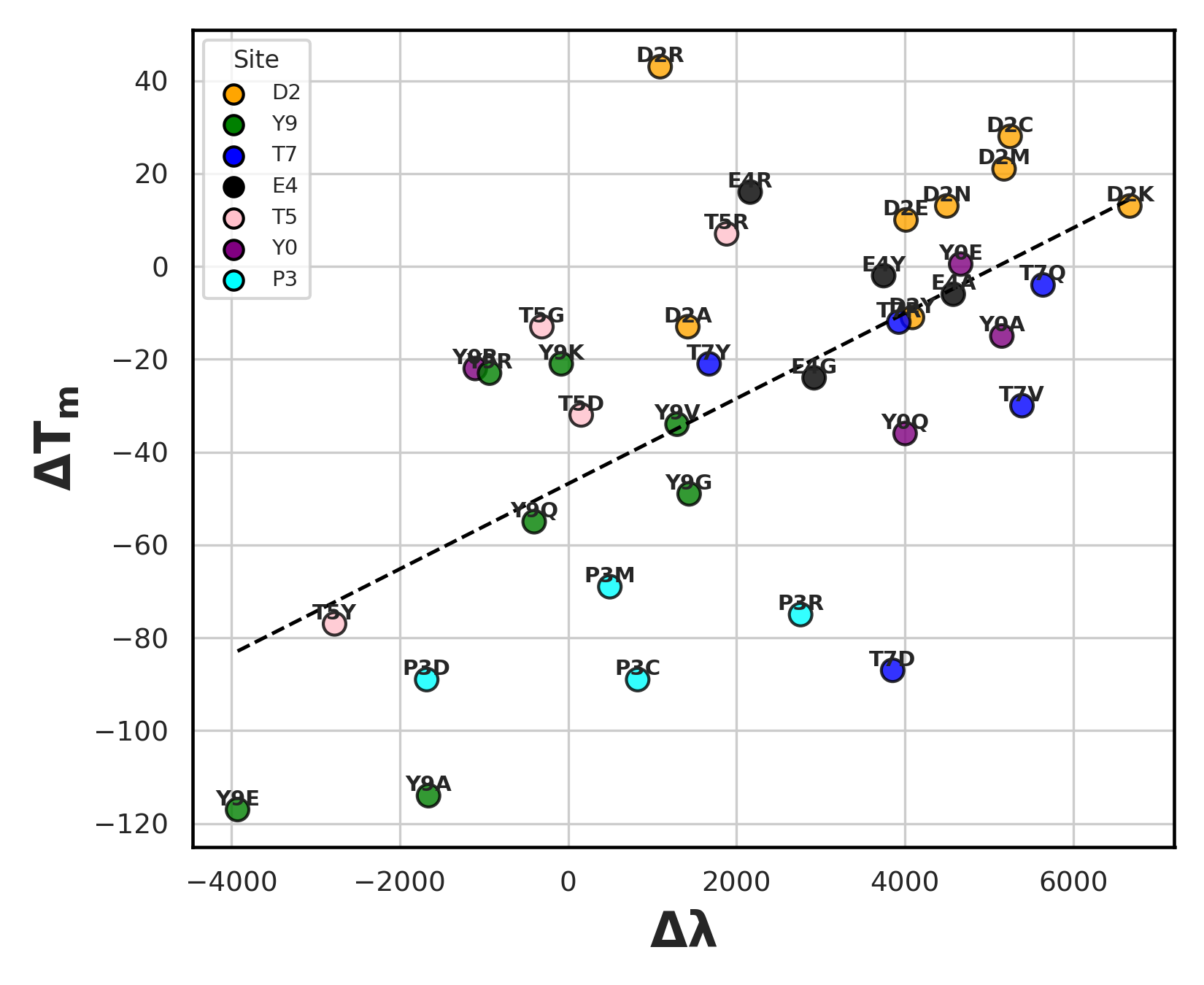}
  \put(2,82){\scriptsize\textbf{(a)}}
\end{overpic}
\label{fig:dhlda_scatter}
\end{subfigure}
\hfill
\begin{subfigure}[b]{0.49\columnwidth}
\centering
\vspace{0pt}
\begin{overpic}[width=\linewidth]{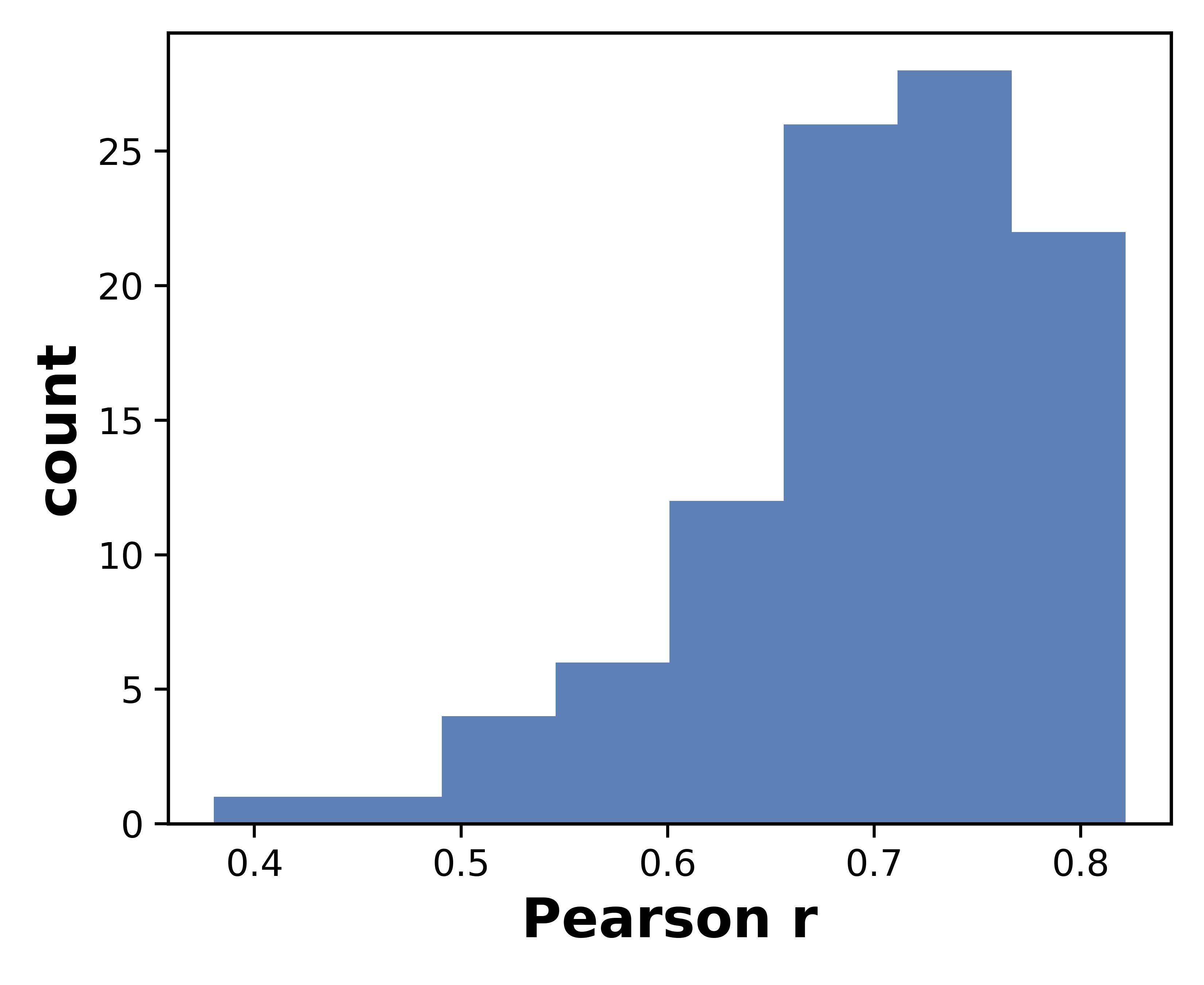}
  \put(2,82){\scriptsize\textbf{(b)}}
\end{overpic}
\label{fig:dhlda_subsampling}
\end{subfigure}
\caption{Mutation-level relationship between HLDA eigenvalue shifts and changes in the peptide thermodynamic conformational stability: (a) Correlation between $\Delta\lambda$ and $\Delta T_m$ (Pearson $r=0.69$, $p=3.4\times10^{-6}$; Spearman $\rho=0.64$, $p=2.3\times10^{-5}$).
(b) Subsampling analysis demonstrating the robustness of the correlation upon excluding a fraction of mutants.}
\label{fig:dhlda_with_subsampling}
\end{figure}

Across the full set of mutants, we observe a positive correlation between $\Delta\lambda$ and the mutation-induced stability changes, quantified by the change in the system's melting temperature $\Delta T_m$ (Fig.~\ref{fig:dhlda_with_subsampling}(a)), suggesting that substitutions which increase folded–unfolded separability along the HLDA coordinate tend to be stabilizing, whereas those that decrease separability tend to be destabilizing. To assess the robustness of this relationship, we performed a subsampling analysis in which the correlation was recomputed 100 times after randomly removing 40\% of the mutants. The resulting distribution of correlation coefficients is centered at  $r=0.75$, indicating that the observed trend is reasonably robust to data subsampling (Fig.~\ref{fig:dhlda_with_subsampling}(b)).

\section*{Discussion and Conclusions}
Tuning the free-energy surface (FES) of biomolecules such as peptides and proteins through point mutations is a highly complex task that is combinatorially intractable. Even for a peptide consisting of only 10 residues, a brute-force approach would require the computation of the FES for $\mathrm{20^{10}}$ possible mutants , making direct computation of mutation-induced free-energy changes impractical. Building on the CV-FEST framework, the goal of this study was to use collective variables(CVs) to guide targeted chemical modifications that systematically reshape the free-energy landscape while avoiding prohibitive computational cost. 

Specifically, we employed HLDA to construct guiding CVs, motivated by its simplicity, interpretability, and ability to be trained using relatively small amounts of data obtained from short unbiased simulations performed within the system’s metastable states. The attained results indicate that this approach can lead to a significant reduction in computational cost. In particular, a CV constructed for the wild-type (WT) system provides residue-level information that appears to be informative of whether replacement of a given residue is more likely to stabilize or destabilize the peptide’s folded state. This behavior is reflected in the observed correspondence between the WT HLDA CV weights and the average change in the peptide melting temperature, as estimated from REMD simulations (Fig.~\ref{fig:wt_res_importance}).

Beyond providing guidance on which residues may be most sensitive to mutation, we further find that the proposed methodology can also assist in informing the choice of amino-acid substitutions at a selected position.  This is achieved by constructing HLDA CVs for individual mutants and comparing the leading HLDA eigenvalue of each mutant to that of the WT system, thereby quantifying changes in the separability of the two basins along the one-dimensional HLDA coordinate. Although this quantity does not represent a direct estimate of the free-energy difference, here too we observe that variations in the leading HLDA eigenvalue are consistent with trends in the mutation-induced shift of the free-energy balance between the basins (Fig.~\ref{fig:dhlda_with_subsampling}). This suggests that the eigenvalue difference may serve as a practical and computationally inexpensive indicator of relative stabilization or destabilization.

Unlike many machine-learning–based approaches to stability prediction that rely on large experimental datasets, a key advantage of the proposed methodology is that it requires no extensive training data. Instead, it relies solely on readily accessible and computationally inexpensive information obtained from short unbiased simulations confined to the metastable states of the systems of interest, thereby capturing intrinsic thermal fluctuations within those states, a concept similarly exploited by interaction-centric tools such as Key Interactions Finder (KIF)\cite{KIF2023}. An important takeaway is that even short trajectories sampled locally within the relevant conformational basins can contain predictive information about the free-energy difference between them, even in the absence of observed transitions. Because the approach is grounded in a physical framework, it may also provide mechanistic insight into the molecular factors governing changes in the FES, a direction which we intend to explore in future work. 

% A similar idea appears in interaction-centric frameworks such as Key Interactions Finder (KIF)~\cite{KIF2023}, which identifies residues and non-covalent interactions associated with user-defined conformational changes using simulation data from separate metastable states. An important implication of this observation is that even short trajectories sampled locally around the relevant conformations can contain predictive information about the free-energy difference between states, despite the absence of any observed transitions during training. Moreover, because the approach is rooted in a physically motivated framework, it may also offer insight into the molecular mechanisms underlying changes in the system’s free-energy surface, an aspect that will be explored in future work. 

As shown, during the development of the methodology, we did observe some dependence on data preprocessing, for example in determining the precise boundaries used to define the states from which the unbiased training data are extracted. While a uniform definition offers a consistent reference across mutants, point mutations may alter state boundaries and transition thresholds, indicating that more system-specific definitions could be beneficial.  In the long term, we envision several possible remedies, including the use of more systematic and automated state-identification procedures that can be adjusted for each mutant, employing alternative descriptors to the RMSD for defining the state boundaries, as well as extending the framework to incorporate more advanced deep-learning architectures for CV construction. In the near term, a practical approach could be to refine state boundaries using a small validation set of mutants for which full REMD calculations are carried out, and then apply the resulting calibrated cutoffs to support efficient exploration of a broader mutation space. To further evaluate the robustness and generality of the approach, future work will extend its application to larger and more complex peptides and proteins.

\section*{Supplementary Information}
The supplementary information includes details on RMSD-based state-boundary selection used to define folded and unfolded ensembles for HLDA (Sec.~S.1), robustness of the HLDA-based correlations to the choice of RMSD thresholds (Sec.~S.2), and REMD convergence diagnostics supporting the reliability of the melting-temperature benchmarks (Sec.~S.3).

\begin{acknowledgments}
The authors acknowledge support from the Israel Science Foundation (ISF) under grant number 1181/24.
\end{acknowledgments}

\section*{Data and Software Availability}

All scripts and input files required to reproduce the analyses in this study are publicly available at
\href{https://github.com/Muralika-M/CV-guided-FES-analysis}{https://github.com/Muralika-M/CV-guided-FES-analysis}.
The repository contains Python scripts for HLDA analysis, threshold scanning, correlation analysis, and figure generation, together with representative example datasets and the GROMACS and PLUMED input files used in this work. Molecular dynamics trajectories are not included due to their size but can be provided by the authors upon reasonable request.

\bibliographystyle{aipnum4-1}
\section*{References}
\bibliography{references}

\clearpage
\setcounter{section}{0}
\renewcommand{\thesection}{S\arabic{section}}
\renewcommand{\thesubsection}{S\arabic{section}.\arabic{subsection}}
\renewcommand{\thefigure}{S\arabic{figure}}
\setcounter{figure}{0}

% Put Supporting Information in single-column layout.
\onecolumngrid

\section*{Supporting Information}

\subsection*{S.1. RMSD-based state-boundary selection}
\begin{figure}[htbp]
  \centering
  \includegraphics[width=0.7\textwidth]{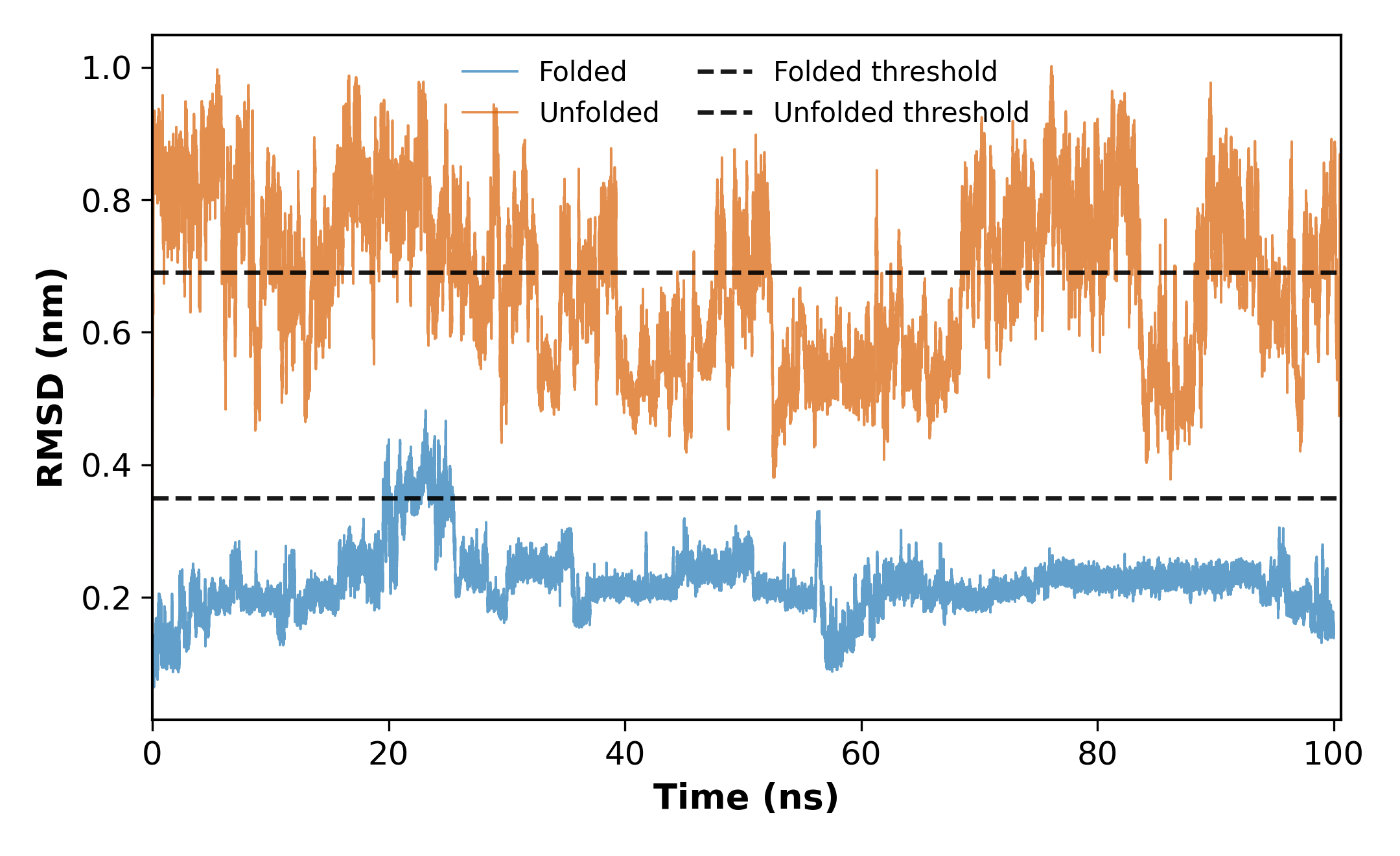}
  \caption{RMSD versus time for the folded- (blue) and unfolded-start (orange) unbiased trajectories. Dashed lines denote the representative RMSD thresholds used here (\(\mathrm{thr}_F=0.35~\mathrm{nm}\), \(\mathrm{thr}_U=0.69~\mathrm{nm}\)), corresponding to the main-text $\Delta\lambda$--$\Delta T_m$ analysis (Fig.~\ref{fig:dhlda_scatter}).}
  \label{fig:rmsd_ub_training}
\end{figure}

Fig.~\ref{fig:rmsd_ub_training} shows RMSD time series for two short unbiased trajectories initialized from folded and unfolded conformations. RMSD was computed relative to the system-specific minimum-enthalpy folded reference structure. Dashed lines indicate a representative folded/unfolded threshold pair used for ensemble labeling (matching the main-text $\Delta\lambda$--$\Delta T_m$ analysis); robustness to threshold choice is evaluated via the threshold scan in Fig.~\ref{fig:threshold_heatmap}.

\subsection*{S.2.RMSD-threshold robustness for HLDA-based correlations}
\begin{figure}[htbp]
  \centering
  \includegraphics[width=0.78\linewidth]{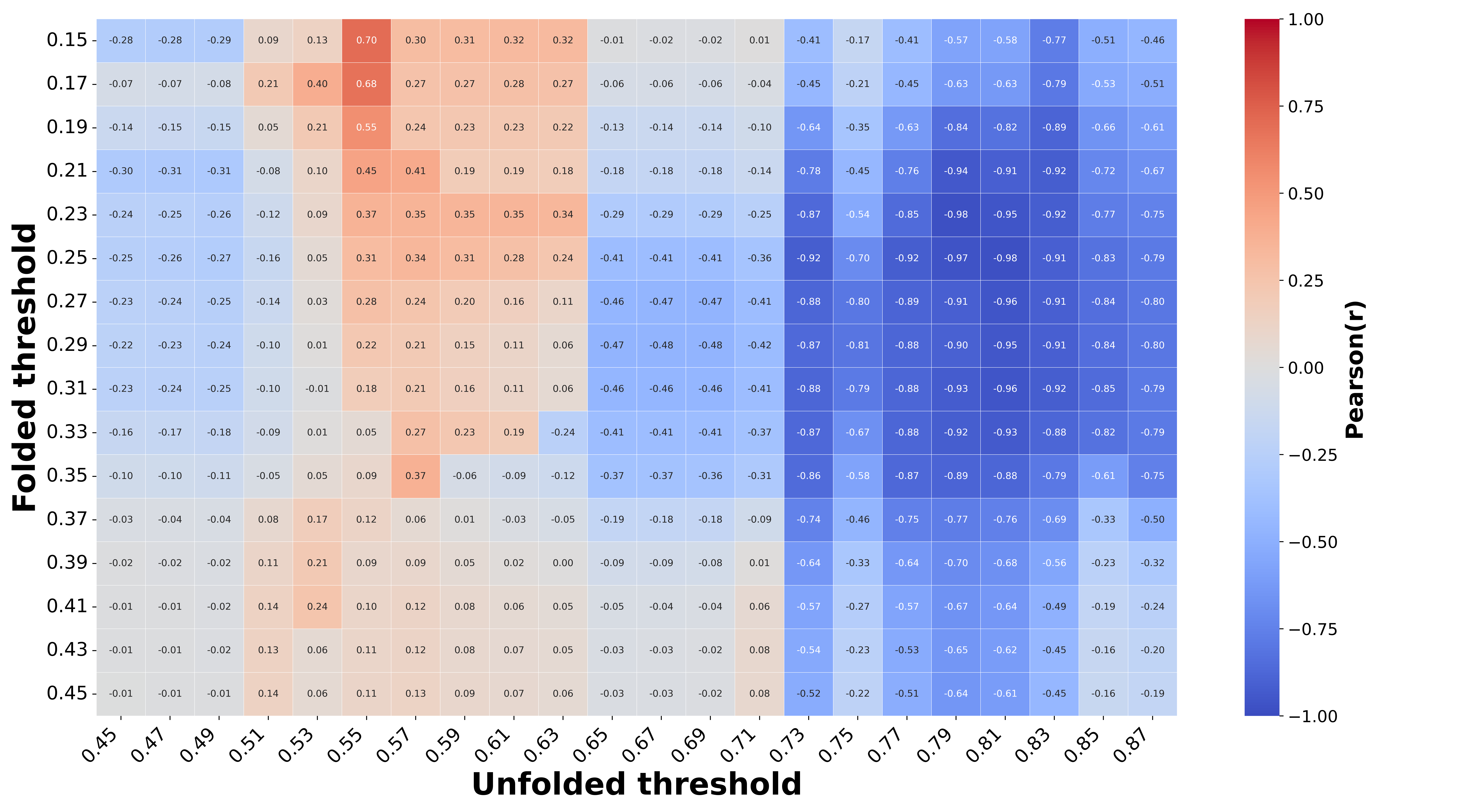}
  \caption{Pearson correlation heatmap between WT HLDA residue importance and average signed $\Delta T_m$ across threshold pairs}
  \label{fig:wt_res_imp_heatmap_si}

  \vspace{1.0cm} 
  
  \includegraphics[width=0.78\linewidth]{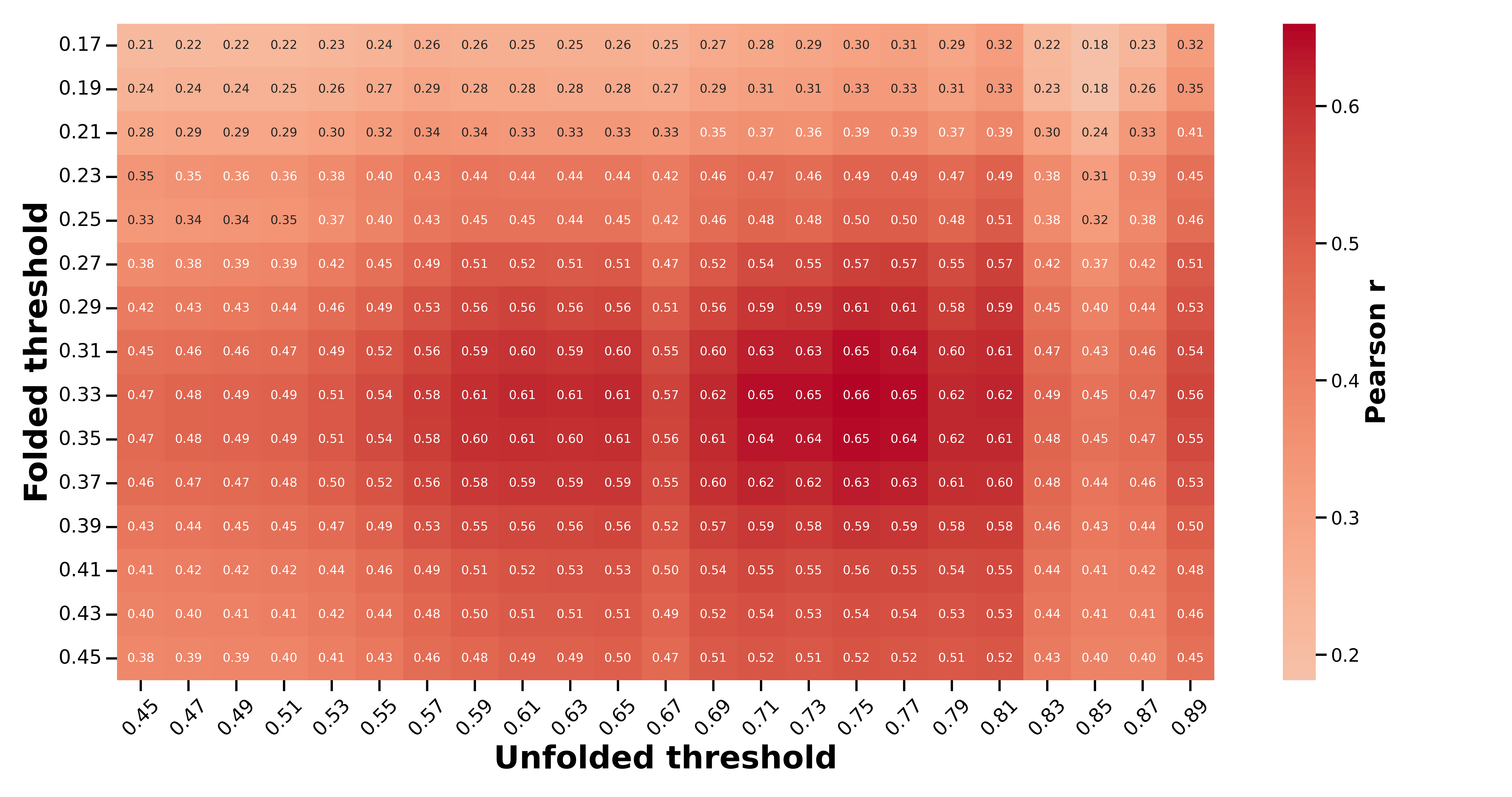}
  \caption{Pearson correlation between $\Delta \lambda$ and $\Delta T_m$ across folded/unfolded RMSD thresholds.}
  \label{fig:threshold_heatmap}
\end{figure}

Figs.~\ref{fig:threshold_heatmap} and \ref{fig:wt_res_imp_heatmap_si} quantify how the reported correlations depend on the RMSD thresholds used to define folded and unfolded ensembles.

\subsection*{S.3. REMD convergence analysis}
\begin{figure}[htbp]
  \centering
  \begin{subfigure}[t]{0.49\columnwidth}
    \centering
    \begin{overpic}[width=\linewidth]{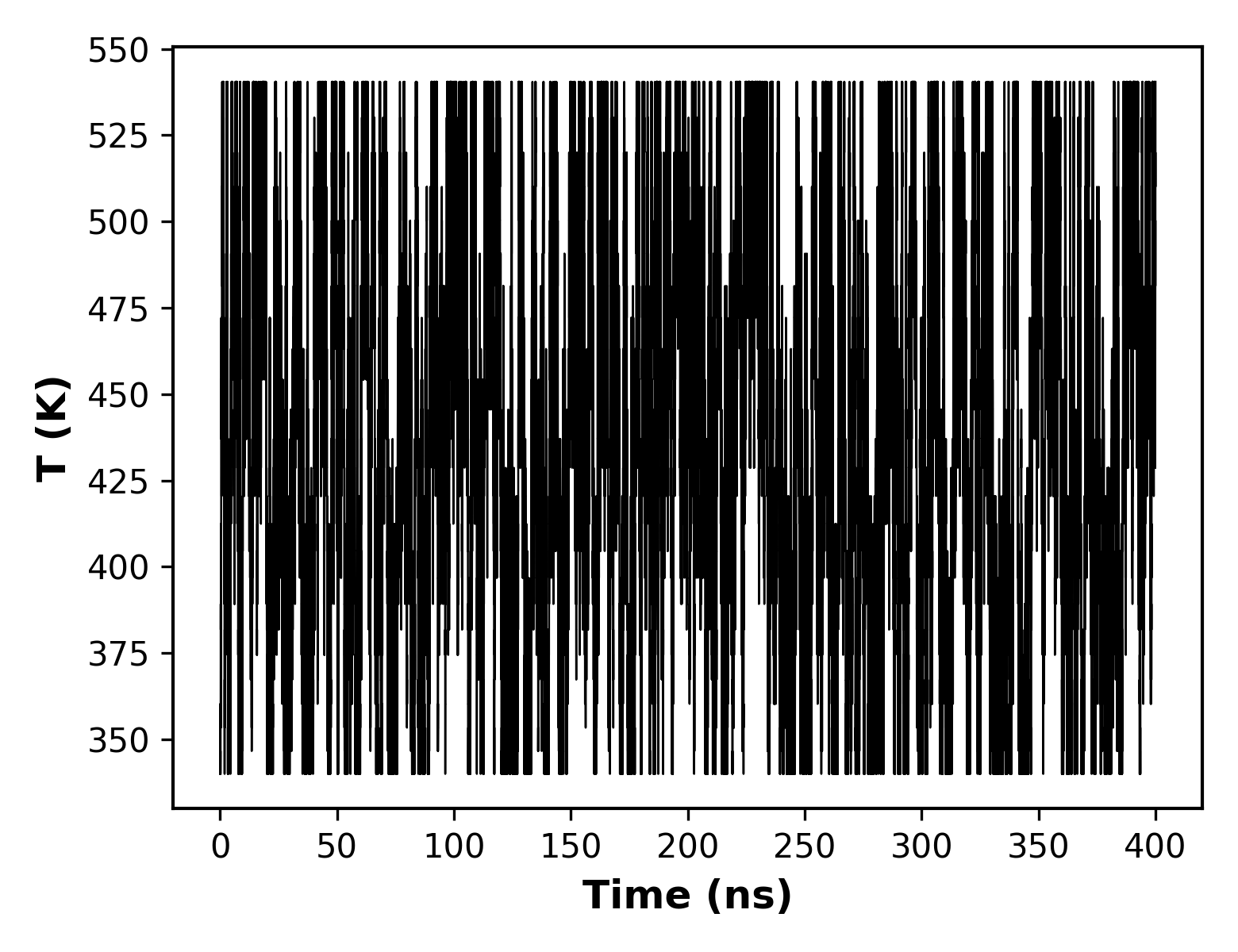}
      \put(1.5,72){\scriptsize\textbf{(a)}}
    \end{overpic}
    \label{fig:si_temp_trace}
  \end{subfigure}
  \hfill
  \begin{subfigure}[t]{0.49\columnwidth}
    \centering
    \begin{overpic}[width=\linewidth]{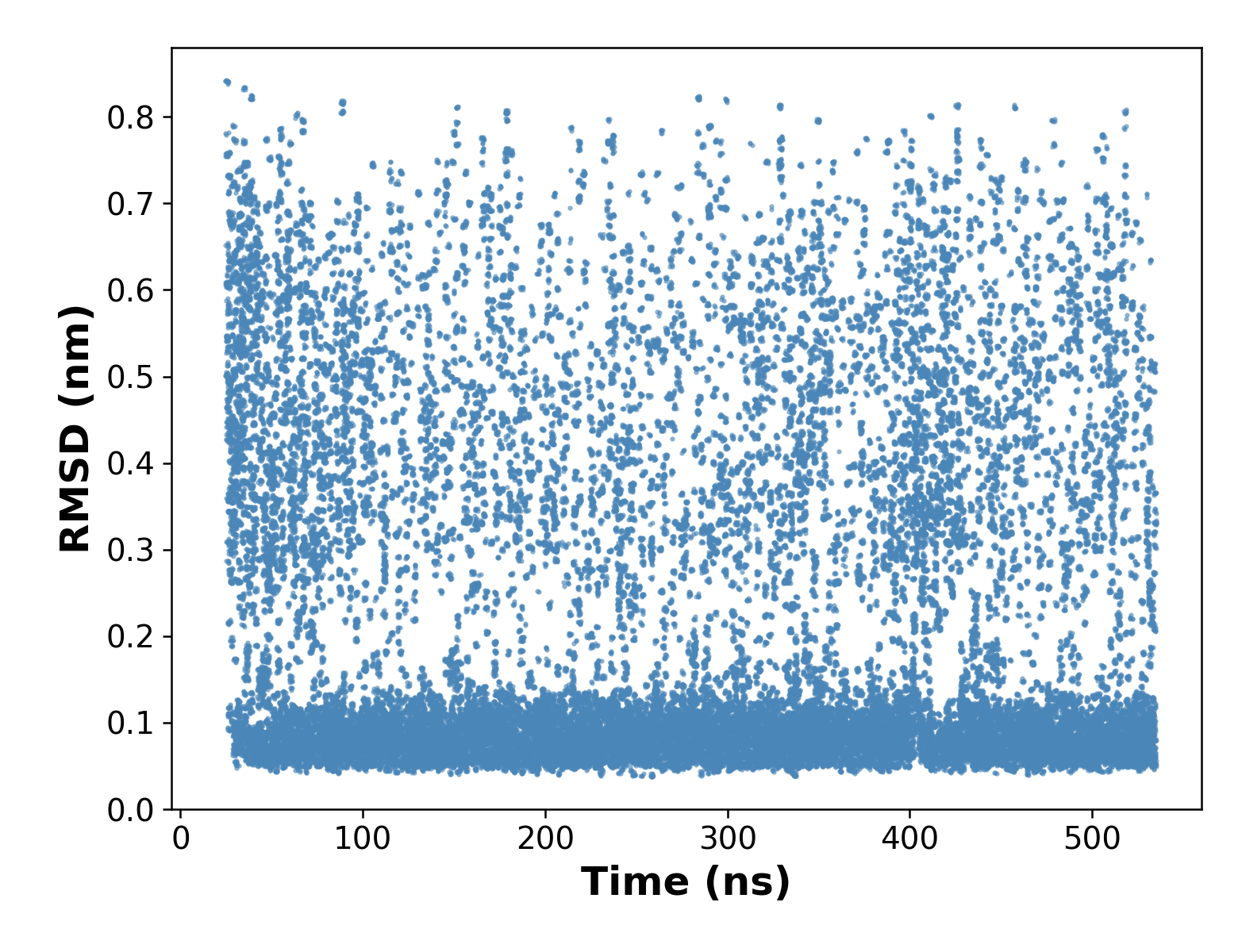}
      \put(2,72){\scriptsize\textbf{(b)}}
    \end{overpic}
    \label{fig:si_rmsd_windows}
  \end{subfigure}
  \vspace{0.5em}
  \begin{subfigure}[t]{0.49\columnwidth}
    \centering
    \begin{overpic}[width=\linewidth]{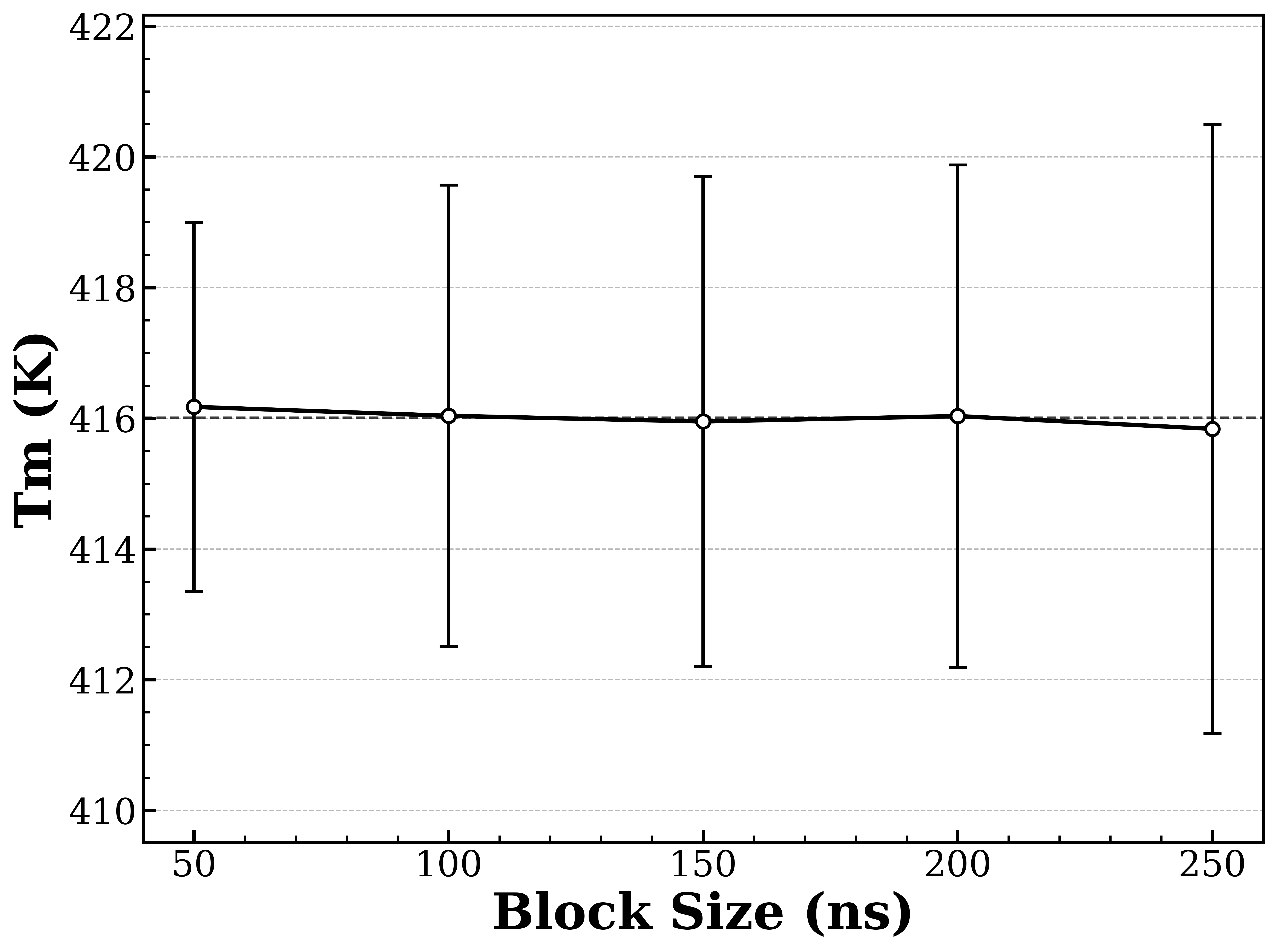}
      \put(1.0,72){\scriptsize\textbf{(c)}}
    \end{overpic}
    \label{fig:si_tm_block_convergence}
  \end{subfigure}
  \caption{REMD convergence diagnostics. (a) Temperature-space trace of the low-\(T\) replica over time. (b) RMSD vs.\ time for the lowest-temperature replica. (c) Block-size convergence of \(T_m\) from multiple independent REMD runs of a system. Points show weighted mean \(T_m\), error bars show weighted SE for block sizes 50--250~ns, and the dashed line indicates the overall mean.}
  \label{fig:si_remd_convergence}
\end{figure}

To assess convergence and reliability of the REMD benchmarks, we evaluated (i) temperature-space mixing, (ii) structural stability over time, and (iii) block-size stability of thermodynamic estimates. Fig.~\ref{fig:si_remd_convergence} shows these diagnostics for a representative system: a temperature trace and RMSD time series for the lowest-temperature replica, together with a block analysis of \(T_m\). The block analysis was performed using multiple independent REMD trajectories, with an aggregate sampling time of more than \(2~\mu\text{s}\). After discarding the first 150 ns of each trajectory as equilibration, the remaining data were partitioned into contiguous blocks of 50, 100, 150, 200, and 250 ns. For each block, \(p_\mathrm{fold}(T)\) was computed across the 25-replica temperature ladder, and \(T_m\) was obtained by interpolation at \(p_\mathrm{fold}=0.5\). Block-wise estimates were then pooled across runs to report a weighted mean and weighted standard error (SE).

\clearpage
\twocolumngrid
\clearpage
\twocolumngrid

\end{document}